\documentclass[5p,sort&compress]{elsarticle}
\usepackage{amssymb}
\usepackage{amsmath}
\usepackage{graphicx}
\usepackage{hyperref}

\usepackage{graphicx}
\usepackage{color}

\usepackage{amsmath}
\usepackage{amsfonts}   
\usepackage{amssymb}  
\usepackage{mathrsfs}
\usepackage{epsfig,bbm}
\usepackage{bm}
\usepackage{dsfont}

\def\laq{~\raise 0.4ex\hbox{$<$}\kern -0.8em\lower 0.62
ex\hbox{$\sim$}~}
\def\gaq{~\raise 0.4ex\hbox{$>$}\kern -0.7em\lower 0.62
ex\hbox{$\sim$}~}

\def\beq{\begin{equation}}
\def\eeq{\end{equation}}
\def\bea{\begin{eqnarray}}
\def\eea{\end{eqnarray}}
\def\bean{\begin{eqnarray*}}
\def\eean{\end{eqnarray*}}

\def \pa {\partial}

\def \ti {\widetilde}

\def \La {\Lambda}

\def \ga {\gamma}
\def \sg {\sigma}

\def \Om {\Omega}

   \def\be{\begin{equation}}
   \def\ee{\end{equation}}
   \def\ba{\begin{eqnarray}}
   \def\ea{\end{eqnarray}}

\def\laq{~\raise 0.4ex\hbox{$<$}\kern -0.8em\lower 0.62ex\hbox{$\sim$}~}
\def\gaq{~\raise 0.4ex\hbox{$>$}\kern -0.7em\lower 0.62ex\hbox{$\sim$}~}

\def\beq{\begin{equation}}
\def\eeq{\end{equation}}
\def\bea{\begin{eqnarray}}
\def\eea{\end{eqnarray}}

\def \pa {\partial}

\def \ti {\widetilde}

\def \La {\Lambda}

\def \ga {\gamma}
\def \sg {\sigma}

\def \Om {\Omega}

    \def\be{\begin{equation}}
    \def\ee{\end{equation}}
    \def\ba{\begin{eqnarray}}
    \def\ea{\end{eqnarray}}

\newcommand{\eq}{\begin{equation}}
\newcommand{\eqx}{\end{equation}}
\newcommand{\eqn}{\begin{eqnarray}}
\newcommand{\eqnx}{\end{eqnarray}}

\newcommand{\Ups}{\Upsilon}

\newcommand{\Hcal}{\mathcal H}

\begin{document}



\title{Time of flight of ultra-relativistic particles in a realistic Universe: \\ a viable tool for fundamental physics?}

\author[fisB,infn,gen]{G. Fanizza}
\ead{Giuseppe.Fanizza@ba.infn.it}
\author[fisB,infn]{M. Gasperini}
\ead{maurizio.gasperini@ba.infn.it}
\author[gen,rio]{G. Marozzi}
\ead{giovanni.marozzi@gmail.com}
\author[coll,cern,sap]{G. Veneziano}
\ead{Gabriele.Veneziano@cern.ch}

\address[fisB]{Dipartimento di Fisica, Universit\`{a} di Bari, Via G. Amendola
173, 70126 Bari, Italy}
\address[infn]{Istituto Nazionale di Fisica Nucleare, Sezione di Bari, Bari, Italy}
\address[gen]{Universit\'e de Gen\`eve, D\'epartement de Physique Th\'eorique and CAP,
24 quai Ernest-Ansermet, CH-1211 Gen\`eve 4, Switzerland}
\address[rio]{Centro Brasileiro de Pesquisas F\'{\i}sicas, Rua
  Dr. Xavier Sigaud 150, Urca,  CEP 22290-180, Rio de Janeiro, Brasil}
\address[coll]{Coll\`ege de France, 11 Place M. Berthelot, 75005 Paris, 
 France}
\address[cern]{CERN, Theory Unit, Physics Department, CH-1211 Geneva 23, Switzerland}
\address[sap]{Dipartimento di Fisica, Universit\`a di Roma La Sapienza, Rome, Italy}

\begin{abstract}

Including the metric fluctuations of a realistic cosmological geometry we reconsider an earlier suggestion that measuring the relative time-of-flight of ultra-relativistic particles can provide interesting constraints  on fundamental cosmological and/or particle parameters. Using convenient properties of the geodetic light-cone coordinates we first compute, to leading order in the Lorentz factor and for  a generic (inhomogeneous, anisotropic) space-time, the relative arrival times of two ultra-relativistic particles as a function of their masses and energies as well as of the details of the large-scale geometry. Remarkably,  the result can be written as an integral over the unperturbed 
line-of-sight of a simple function of the local, inhomogeneous redshift. We then evaluate the irreducible scatter of the expected data-points due to first-order metric perturbations, and discuss, for an ideal source of ultra-relativistic particles, the resulting attainable precision on the determination of different physical parameters.
\end{abstract}



\maketitle

It is well known that times of flight of ultra-relativistic (UR) particles received from a distant astrophysical source depend on the particle mass $m$, on the particle energy $E$ measured by the observer, and on the details of the space-time geometry in which the particle trajectory is embedded. 

The first pioneer study on this subject  \cite{Zat} has shown, in particular, that the observation of the relative arrival times of neutrinos of different energies emitted in Supernovae explosions can provide significant information on neutrino masses. In a later, complementary paper \cite{Stod} it has been pointed out that measuring the relative arrival times of neutrinos and photons (or of different neutrino species), and knowing neutrino masses, energies, and the redshift of the source, one can in principle obtain numerical estimates of cosmological parameters 
(such as the present values of the Hubble and deceleration parameters). 

The results presented in \cite{Zat,Stod} are both based on the homogeneous and isotropic  cosmology described by the standard Friedman-Lema\^itre-Robertson-Walker (FLRW) metric. In this case the 
 flight-time difference between two UR particles, emitted by the same source at time $\tau_s$, can be expressed, to lowest order in the inverse Lorentz factor $\gamma^{-1}= m/E$, as \cite{Stod}:
\beq
\Delta \tau =  \tau_1 - \tau_2 = \left(\frac{m_1^2}{2E_1^2} - \frac{m_2^2}{2E_2^2} \right)  \int_{\tau_s}^{\tau_o} 
\frac{d \tau}{1+z(\tau)} .
\label{1}
\eeq
Here  $\tau$ is the proper time of the observer (with $\tau_o$ the arrival time at the observer of massless 
particles emitted by a source at time $\tau_s$), $m_{1,2}$ and  $E_{1,2} \gg m_{1,2}$ are energies and masses of  the two particles as measured by the observer, and $z$ is the cosmological redshift $1+z=a_o/a$, where $a$ is the scale factor of the FLRW geometry.

The Universe, however, is full of structure at different length scales. An interesting question is how Eq. (\ref{1}) is affected by inhomogeneities when these are not assumed to be negligible. A priori one might expect that inhomogeneities could alter (\ref{1})  by terms proportional to a lower power of $m/E$. More generally, such effects should be taken into account if one wants to connect precisely the data to cosmological and/or particle physics parameters.
In this Letter we  exploit the remarkable properties of the so-called  geodetic light-cone (GLC) coordinates \cite{1} to answer the above questions. The basic simplification is that null geodesics are extremely simple to describe in GLC coordinates. UR (or nearly null) geodesics turn out to be sufficiently simple for the problem to be tractable.

We start by recalling the definition of GLC coordinates \cite{1} and some already well known properties of them (see also \cite{Fleury:2016htl} for a recent discussion). They consist of a timelike  coordinate $\tau$,  a null coordinate $w$, and two angular coordinates $\tilde{\theta}^a$ ($a=1,2$). The parameter $\tau$ can  be  identified with the proper time in the synchronous gauge and thus provides the four-velocity of a  static geodesic observer in the form $u_{\mu} = -\partial_{\mu} \tau$. The GLC line-element  depends on six arbitrary functions ($\Ups, U^a, \ga_{ab}=\ga_{ba} $, $a,b=1,2$), 
and takes the form:
\beq
ds^2\! = \!\Ups^2 dw^2\!-\!2\Ups dw d\tau+ \! \gamma_{ab}(d\tilde\theta^a\!- \! U^a dw)(d\tilde\theta^b\!-U^b dw),
\label{4}
 \eeq
where $\gamma_{ab}$ and its inverse  $\gamma^{ab}$ lower and  raise  the two-dimensional indices. 

In the GLC coordinates the (interior of the) past light-cone of a given observer is defined by  $w = (<)~ w_o =$ constant. Furthermore, null geodesics stay at fixed values of the angular coordinates $\tilde{\theta}^a= \tilde{\theta}^a_o =$ constant, with $\tilde{\theta}^a_o$ specifying the source direction at the observer position. Finally, the redshift $z$ of a signal propagating along a light-cone, emitted at time $\tau$ by a comoving source and received at time $\tau_o$ by a comoving observer, is given by a simple generalization of the standard FLRW expression:
\be
1+z= 
{\Ups(\tau_o, w_o, \ti\theta^a_o)}/{\Ups(\tau, w_o, \ti\theta^a_o)}.
\label{5}
\ee
The above properties of the GLC coordinates have already found several interesting applications \cite{1,Fleury:2016htl,Fanizza:2013doa,11a,3,BenDayan:2012pp,BenDayan:2013gc,Nugier:2013tca,Ben-Dayan:2014swa,Marozzi:2014kua,DiDio:2014lka, Fanizza:2014baa, Fanizza:2015swa}.  
In the present context we are interested in describing a family of almost null geodesics that start from a source lying on a past light-cone $w = w_o$ at a given $z$. The geodesics, however, reach the observer at later values of $w$, $w = w_i$. The latter will depend on the  Lorentz factor $\gamma_i$ of the $\rm{i^{th}}$ particle  which thus travels between the two  light-cones $w=w_o$ and $w = w_i$.

We then write down the standard geodesic equation and mass-shell constraint for a point particle of mass $m$, propagating in  the metric (\ref{4}). The latter condition reads
\beq
\label{C}
2 (\Ups \dot{\tau})\dot{w} - \gamma_{ab} \dot{{\tilde{\theta}}}^a \dot{{\tilde{\theta}}}^b  + 2 U_a \dot{{\tilde{\theta}}}^a \dot{w}  - (\Ups^2 + U^2) \dot{w}^2   = 1 \, ,
\eeq
where a dot denotes differentiation with respect to the particle's proper time $ds = \sqrt{- dx^{\mu} d x ^{\nu} g_{\mu\nu}}$. In order to make the extrapolation to the massless limit smooth, let us rescale proper time by  the Lorentz factor at the observer, $\gamma_o$. In that case the r.h.s. of Eq. (\ref{C}) becomes $m^2 /E^2 \ll1$, with $E$ the energy measured by the observer.
Our claim now is that there is a perturbative hierarchy among the quantities $\dot{\tau}, \dot{w}, \dot{{\tilde{\theta}}}^a$, with:
\beq
\label{hierarchy}
\dot{\tau} \sim \gamma^0,  ~~~~~ \dot{{\tilde{\theta}}}^a  \sim \gamma^{-1}, ~~~~~\dot{w} \sim \gamma^{-2} \, .
\eeq
We will check below that such an assumption is self consistent. Assuming  it, we can rewrite (\ref{C}) in the form:
\beq
\label{C1}
2 (\Ups \dot{\tau})\dot{w} - \gamma_{ab} \dot{{\tilde{\theta}}}^a \dot{{\tilde{\theta}}}^b  + 2 U_a \dot{{\tilde{\theta}}}^a \dot{w}  +\dots = \frac{m^2}{E^2}\, ,
\eeq
where the dots represent next to next to leading contributions.
Analogously, the geodesic equations read:
\bea
\!\!\!\!\!\!\!\!\!\!\!\!(\Ups \dot{\tau})^{.}\!\!&=&\!\!\left( U^a \gamma_{ab, \tau} - U_{b,\tau} \right) \dot{\tau} \dot{{\tilde{\theta}}}^b 
+ \dots \,, 
\label{taueq}  \\ 
 \!\!\!\!\!\!\!\!\!\!\!\!\ddot{w}\!\!&=&\!\!- \frac{1}{2 \Ups} \gamma_{ab, \tau} \dot{{\tilde{\theta}}}^a \dot{{\tilde{\theta}}}^b -   \frac{1}{ \Ups} \left( \Ups_{,a} - U_{a,\tau} \right) \dot{w} \dot{{\tilde{\theta}}}^a + \dots \,,
\label {weq} \\
\!\!\!\!\!\!\!\!\!\!\!\!\ddot{{\tilde{\theta}}}^a\!\!&=&\!\!- \gamma^{ab} \gamma_{bc, \tau}  \dot{\tau} \dot{{\tilde{\theta}}}^c  - \gamma^{ab} \left( \Ups_{,b} - U_{b,\tau} \right)  \dot{\tau} \dot{w}  \nonumber \\ && 
\!\!\!\!\!\!\!\!\!\!\!\!-\left( \gamma^{ab} \Gamma_{cd~b} + \frac{1}{2\Ups} U^a \gamma_{cd, \tau} \right)  \dot{{\tilde{\theta}}}^c \dot{{\tilde{\theta}}}^d + \dots \, ,
\label{thetaeq}
\eea
where $\Gamma_{cd~b} = \frac12(\gamma_{cb,d} +\gamma_{db,c} - \gamma_{cd,b}) $. It is a straightforward (though tedious) exercise to verify that, to next to leading order included, the constraint (\ref{C}) is preserved by the evolution equations (\ref{taueq}), (\ref{weq}) and (\ref{thetaeq}).

At the same level of approximation, we find immediately from (\ref{C1}) that
\beq
\label{wdot}
2 \dot{w} = \frac{\frac{m^2}{E^2}
 + \gamma^{ab} J_aJ_b}{\Ups \dot{\tau} + U_a \dot{{\tilde{\theta}}}^a}\, ,
\eeq
where $J_a \equiv \gamma_{ab} \dot{{\tilde{\theta}}}^b$. This equation is clearly consistent with (\ref{hierarchy}) since the numerator is of order $\gamma^{-2}$ while the denominator is of $O(1)$ with a relative correction  $O(\gamma^{-1})$.
Another straightforward calculation shows that (\ref{wdot}) gives the correct result for $\ddot{w}$ once 
Eqs. (\ref{taueq}-\ref{thetaeq}) are used. A useful imput for this check is the smallness ($O(\gamma^{-2})$) of the first derivative of $J_a$
\beq
\label{Jeq}
\dot{J}_a = \frac12 \left( \gamma_{bc,a} - \frac{1}{\Ups} U_a \gamma_{bc,\tau}\right)  \dot{{\tilde{\theta}}}^b \dot{{\tilde{\theta}}}^c -  \left( \Ups_{,a} - U_{a,\tau} \right) \dot{\tau} \dot{w} \, .
\eeq
 
The quantity we need to compute is  $dw/d\tau = \dot{w}/\dot{\tau}$.  From (\ref{wdot}) we  obtain, to leading order in $m/E$,
 \beq
 \label{dwdtau}
 \frac{dw}{d\tau} =  \frac{\Ups}{2 (\Ups \dot{\tau})^2} \left(\frac{m^2}{E^2}
 + \gamma^{ab} J_aJ_b \right) \, .
 \eeq
 
 We now note that the time dependence of both $\Ups \dot{\tau}$ and $J_a$ appears only at higher order in $m/E$ thanks to Eqs. (\ref{taueq}) and (\ref{Jeq}), respectively. Evaluating $\Ups \dot{\tau}$ at the observer gives simply $\Ups_o\equiv \Ups(\tau_o, w_o, {\tilde{\theta}}_o)$ (because of the rescaling we adopted on the proper time). Integrating now (\ref{dwdtau}) from the source to the observer (along the geodesic) gives:
 \beq
 \label{deltaw}
w_i - w_o =  \frac12 \int_{\tau_s}^{\tau_o} d\tau \frac{\Ups}{\Ups_o^2} (\frac{m_i^2}{E_i^2}
 + \gamma^{ab} J_aJ_b)\,,
 \eeq
 where, to this order in $ \gamma^{-1}$, $\tau_i$ has been taken equal to $\tau_o$.
 There are two further simplifications that we can apply to our final result (\ref{deltaw}). The first is that $J_a$  is zero at the observer (and then also all along the geodesic, because of its approximate constancy) for a geodesic arriving exactly at the observer. The same is true for the quantity $ \gamma^{ab} J_aJ_b$ appearing in (\ref{deltaw}), since it can also be written as $\gamma_{ab} \dot{{\tilde{\theta}}}^a \dot{{\tilde{\theta}}}^b$. 
 The second observation is that the integral in Eq. (\ref{deltaw}) can be taken along the unperturbed null geodesic (with constant ${\tilde{\theta}}^a$ and $w$), since deviations from it are subleading.
 
 Let us finally, compare two such geodesics starting from the same source at the same time $\tau_s$. Their relative time delay can be easily obtained by subtracting two equations like (\ref{deltaw}) to yield:
 \bea
 \label{final}
\! \!\!\!\!\!\!\!w_{1} - w_{2} \!&=&\!  \left(\frac{m_1^2}{2E_1^2} - \frac{m_2^2}{2E_2^2} \right)
\int_{\tau_s}^{\tau_o} d\tau \frac{\Ups}{\Ups_o^2} (\tau, w_o, {\tilde{\theta}}^a_o) \,,\nonumber \\
\!\!\!\!\!\!\!\!\tau_1 - \tau_2 \!&=&\!  \left(\frac{m_1^2}{2E_1^2} - \frac{m_2^2}{2E_2^2} \right) \int_{\tau_s}^{\tau_o} \frac{d\tau}{1
+z(\tau, w_o, {\tilde{\theta}}^a_o)} \,,
 \eea
 where we used Eq. (\ref{5}) and $ \Delta \tau \equiv \tau_1 - \tau_2 = \Ups_o( w_{1} - w_{2})$ (see also \cite{1}). 
 
This is our main result showing that, to leading order in $\gamma_{1,2}^{-1}$, the arrival-time  difference is very similar to the FLRW expression in Eq. (\ref{1}),
 with the only difference that the redshift along the (massless) line-of-sight, being the exact redshift associated with a generic (inhomogeneous and anisotropic) geometry, is  no longer just a function of time. The obtained geometric corrections, 
to leading order again, are the same for the two particles and thus factor out in the time-delay difference. The simplicity of (\ref{final}) suggests that a general derivation of it should exist, even without using a particular coordinate system as done here.

The result (\ref{final}), beside being intrinsically important for the study of UR geodesics, allows us to check the interesting possibility  that future measurements of time-of-flight differences $\Delta \tau$ may provide significant physical information  even after taking into account  the dispersion due to large-scale inhomogeneities.
Let us compute, to this purpose, the irreducible statistical error for a single $\Delta \tau$ measurement, due to  a realistic stochastic background of scalar metric perturbations.
These are described -- to first order and in the absence of anisotropic stresses -- by the Bardeen potential $\psi$ which appears in the metric of the Poisson (or longitudinal) gauge (PG) as \cite{Muk}:
\beq
 ds^2\! = \! a^2(\eta) \! \left[\!  - \! d\eta^2(1\! +\! 2\psi) \! +\! (1\! -\! 2\psi) (dr^2+ r^2 d\Om^2)\right].
 \label{15}
\eeq
We begin by expressing the value of $\Delta \tau$ as a function of $\psi$ and of the observational coordinates (the redshift of the source and the direction of observation). 
For this, we need the first order expansion of $\Ups(\tau, w_o, {\tilde{\theta}}_o)$ in term of
$\psi$, and a further expansion of $\Ups$ and 
of the lower limit $\tau_s$ of the integral in (\ref{final}) around
a time coordinate  
$\bar{\tau}^{(0)}$  such  that the observed redshift $z$ is given by $1+z=\Ups(\tau_o, w_o, {\tilde{\theta}}_o)/\Ups(\tau, w_o, {\tilde{\theta}}_o)= a_o/a( \bar{\tau}^{(0)})$.
We then obtain (see \cite{Fanizza:2015swa} for details):
\bea
\Ups &=& \Ups^{(0)}(\bar{\tau}^{(0)})+\Ups^{(1)}(\bar{\tau}^{(0)}, w_o, {\tilde{\theta}}^a_o)+ \cdots \,, \nonumber \\
\tau_s &=& \bar{\tau}_s^{(0)}+ \bar{\tau}_s^{(1)}(\bar{\tau}_s^{(0)}, w_o, {\tilde{\theta}}^a_o)+ \cdots \,,
\label{FirstExp}
\eea
where $\Ups^{(0)}$ coincides with the scale factor of the unperturbed metric in the PG.

Hereafter, for simplicity, we shall explicitly omit two types of perturbative corrections: those describing the effects of peculiar  velocity and those arising from the Bardeen potential both evaluated at the observer position.
The reason is that the uncertainty due to the first type of terms can been effectively removed, as usual, by taking into account the actual value of the observer's proper velocity inferred from observational data (e.g, from CMB dipole measurements). The second type of terms, on the other hand, only gives a sub-leading contribution to the variance (i.e. to the dispersion of the single experimental points around their mean value) that we are going to estimate in this Letter for the observable $\Delta \tau$ (for this second point, see, in particular, the discussion after Eq. (\ref{21})). 

Using the results for $\Ups^{(1)}$ and $\bar{\tau}_s^{(1)}$ previously obtained in \cite{Fanizza:2015swa}, 
we find that the perturbed value of the flight-time difference (\ref{final}) can then be written as 
\begin{equation}
\Delta \tau=\tau_1-\tau_2=\Delta \tau^{(0)} (1+\delta \tau^{(1)})\,,
\label{DeltaPert}
\end{equation}
where $\Delta \tau^{(0)}$ is the unperturbed result in Eq.(\ref{1}),
while  $\delta \tau^{(1)}$  is the fractional correction due to the first-order scalar perturbations, and is given by:
\begin{eqnarray}
&&
\!\!\!\!\!\!\!\!\!\!\!\!\!\!\!\!\!\!\!\!\!\!\delta \tau^{(1)}=\left[
\int_{\bar{\eta}_s^{(0)}}^{\eta_o}d\eta \frac{a^2(\eta)}{a_o}\right]^{-1} 
\int_{\bar{\eta}_s^{(0)}}^{\eta_o}d\eta \frac{a^2(\eta)}{a_o}\nonumber \\
& & \! \!
\times \bigg\{
\left( \frac{\pa_{\eta}\Hcal}{\Hcal^2}-1\right)   \left[
\psi+2\,\int_{\eta}^{\eta_o}d\eta'\pa_{\eta'}\psi(\eta')+v_\rVert
\right]\bigg.\nonumber\\
& &\bigg. \! \!\! \!
+\!\frac{1}{\Hcal(\eta)}\left[
\pa_{\eta}\psi+\pa_r v_\rVert
\right] +\psi(\eta) 
\bigg\}\,.
\label{LG}
\end{eqnarray}
In the above equations
$\psi(\eta)$ stands for $\psi(\eta, \eta_o-\eta, \tilde{\theta}^a_o)$. Also, we have  defined $\Hcal(\eta)= \pa_\eta (\ln a)$, and we have called $\bar{\eta}_s^{(0)}$ the conformal time parameter related to $\bar{\tau}_s^{(0)}$ and to the observed redshift $z_s$ by 
$d \bar{\eta}_s^{(0)}= d \bar{\tau}_s^{(0)}/a=- (1+z_s)^{-1}\Hcal^{-1} dz_s$. 
Finally, we have  introduced the so-called velocity perturbation defined by 
$$
v_\rVert (\eta)= \frac{1}{a(\eta)}\int_{\eta_\text{in}}^{\eta} d\eta' a(\eta')\pa_r\psi(\eta',\eta_o-\eta,{\tilde{\theta}}_o^a)\,,
$$ 
where $\eta_\text{in}$ denotes an early enough initial time when the integrand was negligible 
\cite{BenDayan:2012pp}.

We are now in the position of evaluating the variance of $\Delta \tau/\Delta \tau^{(0)} $, i.e. the quantity controlling the dispersion of its experimental value around its mean value, due to the inhomogeneity corrections given in Eq. (\ref{LG}). Following \cite{BenDayan:2012pp}, and assuming that  $\psi$ describes a stochastic background of metric fluctuations with vanishing statistical (or {\em ensemble}) average ($\overline \psi=0$), 
we can obtain the intrinsic  dispersion of the flight-time measurements over its background value (let us call it $\sigma$) as:
\beq 
\sigma=\sqrt{\overline{\langle (\delta\tau^{(1)})^2 \rangle}}.
 \label{Disperion-DeltaTau}
\eeq
The brackets $\langle\cdots\rangle$  denote angular average, the overbar $\overline{\cdots}$ statistical average (the computation of the mean value  $\overline{\langle \Delta \tau/\Delta \tau^{(0)} \rangle}$ requires the inclusion of second order perturbations, and is postponed to future work).

In order to compute $\sigma$ it is now convenient to decompose the Bardeen potential into Fourier modes:
\beq
\psi(\eta, \vec x)=\frac{1}{\left( 2\pi \right)^{3/2}}\int d^3k\,e^{i \vec k \cdot \vec x}\psi_k(\eta)E(\vec k),
\label{21}
\eeq
where $E$ is a unit statistical variable satisfying the conditions $E^\star(\vec k)= E(-\vec k)$, $\overline{E(\vec k)}=0$ and $\overline{ E(\vec k)E(\vec k') }=\delta(\vec k+\vec k')$. 
We then insert the above expansion into Eqs. (\ref{LG}) and (\ref{Disperion-DeltaTau}), and, assuming statistical isotropy ($\psi_k$ only dependent on $|\vec k|$), we can easily perform the angular integration.
Proceeding in this way we can convince ourselves that, at least for the range of $z_s$ discussed in this paper, the leading 
contribution\footnote{Let us recall here that the additional (possibly large) contributions to $\sg$ related to the intrinsic velocity of the observer have been subtracted from Eq. (\ref{LG}), by taking into account the observational value of such velocity that can be inferred from current experimental data.} to $\sigma$ comes from the term containing the so-called redshift space distortion ($\pa_r v_\rVert$),  being the term with the highest power of $k$ in Fourier space (see e.g. the discussion in \cite{11a,BenDayan:2013gc}).

Keeping only this leading contribution, and considering the linear regime where the time evolution of the perturbation modes can be appropriately parametrized by the growth factor $g(\eta)$ as 
$\psi_k(\eta)= g(\eta) \psi_k(\eta_o)/g(\eta_o)$, we can write down our final estimate for the expected dispersion as follows:
\begin{eqnarray}
& &\!\!\!\!\!\!\!\! \!\!\!\! \!\!\!\!  \!\!\!\!  \sigma^2 \simeq
\overline{\langle (\delta\tau^{(1)}_{\rm lead})^2 \rangle}=
\left[ \int_{\bar\eta_s^{(0)}}^{\eta_o} d\eta\frac{a^2(\eta)}{a^2_o}  \right]^{-2}\!\!
\int_0^{\infty}\!\!\!dk\,k^3{\cal P}(k,\eta_o)\nonumber\\
&& \!\!\!\!\!\!\!\! \!\!\!\! \!\!\!\! \!\!\!\! 
\times \!\!\!\int_{\bar\eta_s^{(0)}}^{\eta_o}\!d\eta \frac{a^2(\eta)f(\eta)}{a^2_o\Hcal(\eta)} 
\!\!\int_{\bar\eta_s^{(0)}}^{\eta_o}\!\!d\eta' \frac{a^2(\eta')f(\eta')}{a^2_o\Hcal(\eta')} {\mathcal I}\left(k (\eta - \eta' )\right) ,
\label{dispersion}
\end{eqnarray}
where $f(\eta)=\int_{\eta_\text{in}}^{\eta}d\eta' [{a(\eta')}/{a(\eta)}][{g(\eta')}/{g(\eta_o)}]$,
\beq
 {\mathcal I}(x) = \frac{\sin x}{x}\left(1- \frac{12}{x^2} + \frac{24}{x^4} \right) +  \frac{4 \cos x}{x^2}\left(1- \frac{6}{x^2}  \right), \nonumber
\eeq
and we have used the dimensionless power spectrum,\\ ${\cal P}(k,\eta)=k^3 |\psi_k(\eta)|^2/(2 \pi^2)$.

For a realistic estimate of the expected dispersion we shall now consider a spectrum of scalar perturbations of inflationary origin, parametrized by an amplitude $A$, a spectral index $n_s$ and a pivot scale $k_0$. Following the standard conventions (see e.g. \cite{Dur})
we set 
${\cal P}(k,\eta)= (9/25)(g(\eta)/g_\infty)^2 A(k/k_0)^{n_s-1}T^2(k)$, where $T(k)$ is the so-called transfer function which takes into account the sub-horizon evolution of modes re-entering during the radiation era, 
and $g/g_\infty$ is well approximated by a simple function of the critical density parameters  (see, for example, \cite{Eisenstein:1997ik}) with $g_\infty$ normalized in such a way that  $g(\eta_o)=1$.
We will use for $T(k)$ the parametrization presented in \cite{Eisenstein:1997ik}, and for the primordial spectrum the approximated values $A= 2.2 \times 10^{-9}$, $n_s= 0.96$, $k_0= 0.05\,{\rm Mpc}^{-1}$, as reported in \cite{pdg}.

As a first, illustrative example we consider the standard CDM-dominated model, where $g(\eta)\equiv1$
and $(a/a_o)=(\eta/\eta_o)^2$. In that case all time integrations can be performed analytically, and we have explicitly checked that the perturbative contributions appearing in Eq. (\ref{LG}), and not included into our expression (\ref{dispersion}), are indeed subleading and provide negligible corrections to our estimate of the dispersion. 
We have numerically integrated over the Fourier modes using the transfer function of \cite{Eisenstein:1997ik} with $k_{\rm eq}= 0.07 h^2 \Om_{m0}\,{\rm Mpc}^{-1}$, $h=0.67$, $a_o=1$ and $\Om_{m0}=1$ (without including baryon effects); also, we have integrated between the infrared cutoff given by the present horizon scale $a_oH_0$ and the UV cutoff $k_{UV}= 0.1 h\,{\rm Mpc}^{-1}\simeq 300\, a_oH_0$. 
The obtained result for $\sigma$ is illustrated in Fig. 1 (dotted curve labelled as $\sg_{\rm CDM}$) as a function of $z_s$. 

\begin{figure}[t]
\includegraphics[width=\columnwidth]{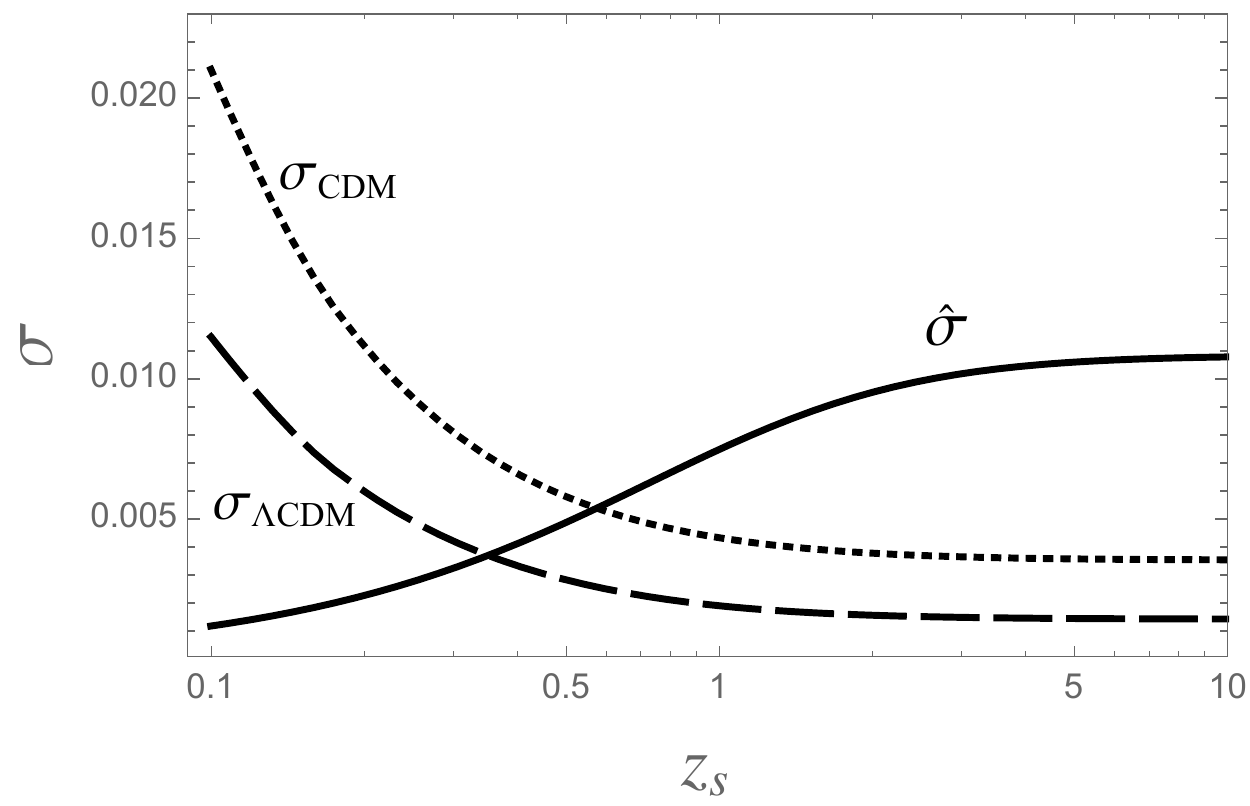}
\caption{
Expected fractional dispersion of measurements of flight-time differences of UR particles, according to Eq. (\ref{dispersion}), as induced by a realistic spectrum of  fluctuations in the CDM (dotted curve) and $\La$CDM  (dashed curve) case. We also plot the fractional dispersion $\widehat\sg$ defined in Eq. (\ref{23}) and associated with the present experimental uncertainty on the value of the cosmological constant (solid curve).}
\label{f1}
\end{figure}

In the case of a background $\La$CDM geometry the spectrum  becomes time dependent and, in
that case, also the time integrals 
in Eq.(\ref{dispersion})
must be performed numerically. 
 Using $\Om_{\La 0}=0.685$  \cite{pdg}, and neglecting again baryons, we obtain  for $\sg$ the result illustrated in Fig. 1 (dashed curve labelled as $\sg_{\La\rm CDM}$).
 
 As before, we have used a UV cut-off  $k_{UV}= 0.1 h\,{\rm Mpc}^{-1}$,
 as the limiting momentum scale below which perturbations can be described in the linear regime.
 To be more realistic, and take into consideration the effect of the mildly non-linear regime, we should include the baryon contribution and use,
  for example,  the so-called HaloFit model
 \cite{Smith:2002dz,Takahashi}
 to describe the non linear evolution of the spectrum.
 On the other hand, our results are rather  insensitive to the choice of the UV cutoff. For instance, by  going from  $k_{UV}= 0.1 h\,{\rm Mpc}^{-1}$ to  
 $k_{UV}= 20 h\,{\rm Mpc}^{-1}$, and including both non-linear
 and baryon effects, would increase the value of $\sigma$ by only about 10$\%$. 
This very weak dependence on the cutoff can be understood from the results obtained in \cite{BenDayan:2013gc} for the averaged  flux.  In fact, 
by  looking at the exact result in CDM,  it can be shown that,
after performing the $\eta$ and $\eta'$ integrations,
 our leading effect for  
 $\sigma$ goes like  $\int dk \,k {\cal P}(k)$, exactly like the case of the average of the flux performed on a sphere of constant redshift embedded in our past light-cone \cite{BenDayan:2013gc}.

To better quantify the possible impact that the scatter of the data illustrated in Fig. 1 might have on future precision measurements of cosmological parameters, let us consider 
the dependence of $\Delta \tau$ from the present value of the 
cosmological constant $\Om_{\La 0}$ by also plotting in Fig. \ref{f1}
(with a solid curve) the dispersion $\widehat\sg$ caused by  the present experimental uncertainty 
$\Om_{\La 0}=0.685^{+0.017}_{-0.016}$. Namely we plot, for an FLRW Universe: 
\be 
\widehat{\sigma} =  \frac{\Delta \tau^{(0)}|_{\Om_{\La 0}=0.669}
-\Delta \tau^{(0)}|_{\Om_{\La 0}=0.702}}{2 \Delta \tau^{(0)}|_{\Om_{\La 0}=0.685}}\,.
\label{23}
\ee 
As clearly shown by the dashed and solid curves of Fig. \ref{f1},
for a source with $z_s \gaq 1$ the expected dispersion due to metric perturbations does not prevent, in principle, the possibility of using UR test particles for measuring cosmological parameters to a   significantly better level of accuracy than the current one. 
On the other hand, the intrinsic dispersion ${\sigma}$ for the $\Lambda$CDM case and the dispersion $\widehat{\sigma}$ due to $\Om_{\La 0}$ are 
nearly the same 
at a redshift $z^* \simeq 0.36$. 
This suggests that we can (or cannot) determinate the value of $\Om_{\La 0}$ with better accuracy than the current one using a single measurement of the time delay if we have (or don't have) 
a suitable source at $z_s > z^*$.

It should be noted that the dispersions $\sg$ and $\widehat{\sigma}$ are both independent on the ratios $m_i/E_i$ corresponding to the two  UR particles. This follows from the non-trivial fact that, to leading order, the geometric corrections associated with large scale inhomogeneities are the same for the two particles and thus factor out in the time-delay difference of Eq. (\ref{final}) (just like the  dependence on the FLRW geometry in Eq. (\ref{1})). 

This factorization  has also the following important consequence.
Consider, for instance, the arrival time of photons and neutrinos emitted  (almost) simultaneously by a distant source at a given redshift $z_s$. In a given FLRW background we can directly connect the 
flight-time difference  to the inverse Lorentz factor $\gamma^{-1}=m/E$ of the neutrinos, according to Eq. (\ref{1}).
However, the large-scale inhomogeneities induce a theoretical error  given by  (\ref{Disperion-DeltaTau}). Therefore,  inferring the  Lorentz factor of the neutrinos 
(or their masses if we measure their energy) from  $\Delta \tau^{(0)}$ will have to take into account this uncertainty. Clearly,
 the fractional error induced on the Lorentz factor will be 
 \be
 {\Delta \gamma}/{\gamma }=  \sigma/2 \,.
 \ee
 
\begin{figure}[t]
\includegraphics[width=\columnwidth]{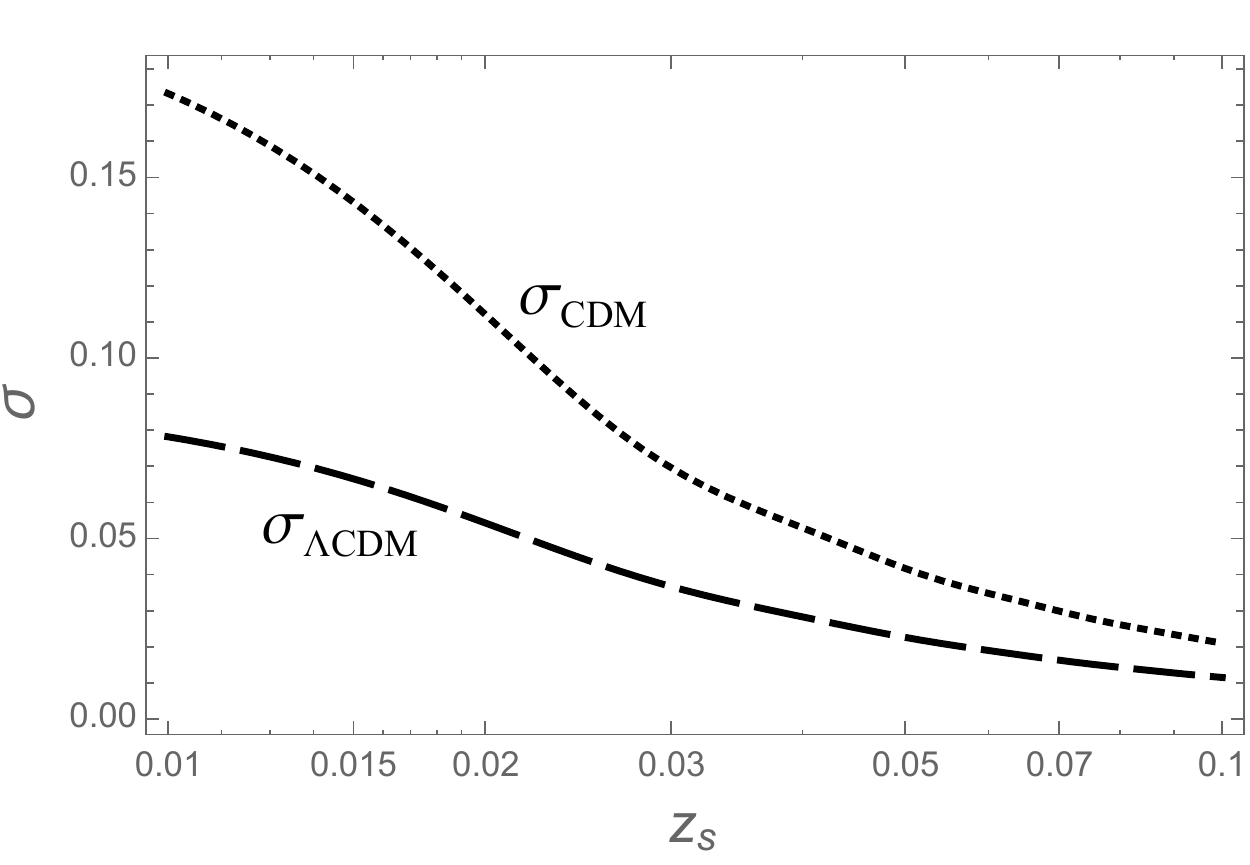}
\caption{Same as Fig. 1 for the range $0.01 < z_s < 0.1$. The dispersion $\widehat\sg$ (not shown)  is negligibly small in this range.} 
\label{f2}
\end{figure}

In Fig. 2 we plot the dispersion $\sigma$ for the range $0.01 < z_s < 0.1$ (which is particularly interesting for the detection of UR particles coming from SNe explosions).
In both the CDM and $\Lambda$CDM case $\sigma$  appears to stabilize at very small $z_s$ at values  about 10\% above those reached at $z_s = 0.01$ (although non-linear effects could enter at very small $z_s$). Picking the $\Lambda$CDM case, we would conclude that the  error induced by inhomogeneities on the determination of the Lorentz factor via flight-time differences should not exceed  5\%.

In conclusion, realistic inhomogeneities do not appear to hamper the possibility of extracting important information on either cosmological or particle physics parameters. Unfortunately, as already pointed out in \cite{Stod}, the main obstacle to making practical use of these ideas remains  the difficulty in finding appropriate sources for which time-of-flight observations are both possible and sufficiently free from other systematic errors.

\section*{Acknowledgements}
GV wishes to thank Leo Stodolsky  for having informed him of his work on the subject and for having asked the question we have tried to answer in this paper. MG wishes to thank Luca Amendola for a clarifying discussion on the cosmic variance. We are also grateful to Fabien Nugier  for interesting comments and suggestions. GF and MG are supported in part by MIUR, under grant no. 2012CPPYP7 (PRIN 2012), and by INFN, under the program TAsP (Theoretical Astroparticle Physics).  GM wishes to thank the Swiss National Science Foundation and CNPq for financial support.

\end{document}